\documentclass[preprint,review,3p,number,prl]{revtex4-1}
\usepackage{newtxtext,newtxmath,amsmath}                       % Permille=\textperthousand

\usepackage{amssymb}
\usepackage{graphicx}
\usepackage{booktabs}
\usepackage{footnote}
\usepackage{tablefootnote}
\usepackage{float}
\usepackage{dcolumn}% Align table columns on decimal point
\usepackage{bm}% bold math
\usepackage[table]{xcolor}% http://ctan.org/pkg/xcolor
\newcommand{\eq}[1]{Eq.~(\ref{#1})}
\newcommand{\Fig}[1]{Fig.~\ref{#1}}
\newcommand{\Figure}[1]{Figure~\ref{#1}}
\newcommand{\tab}[1]{Table~\ref{#1}}

\def\eex#1{$ \times 10^{#1}$}

\def\enhk{\epsilon_{k}}

\def\flux_cm{cm$^{-2}$s$^{-1}$}
\def\flux{m$^{-2}$s$^{-1}$}

\def\pot#1{$^{#1}$}
\def\beqn{\begin{eqnarray}}
\def\eeqn{\end{eqnarray}}

\def\@author#1{\g@addto@macro\elsauthors{\normalsize%
    \def\baselinestretch{1}%
    \upshape\authorsep#1\unskip\textsuperscript{%
      \ifx\@fnmark\@empty\else\unskip\sep\@fnmark\let\sep=,\fi
      \ifx\@corref\@empty\else\unskip\sep\@corref\let\sep=,\fi
      }%
    \def\authorsep{\unskip,\space}%
    \global\let\@fnmark\@empty
    \global\let\@corref\@empty  %% Added
    \global\let\sep\@empty}%
    \@eadauthor={#1}
}

\begin{document}

\title{Corrections to sink strengths used in rate equation simulations of defects in solids}

%\author{T. Ahlgren\corref{cor1}}
%\ead{tommy.ahlgren@helsinki.fi}
%\cortext[cor1]{Corresponding author. Department of Physics, University of Helsinki, P.O. Box 43, %FI-00014 Helsinki, Finland}

\author{T. Ahlgren}
\author{K. Heinola}
\affiliation{Department of Physics, University of Helsinki, P.O. Box 43, FI-00014 Helsinki, Finland}

%\date{\today}

\begin{abstract}

The mean-field rate equations have proven to be a versatile method in simulating defect dynamics and temporal changes in the micro-structure of materials.
However, the reliability and usefulness of the method depends critically on the defect interaction parameters used.
In this study, we show that the sink strength depends also on the detrapping or dissociation process.
%For defects that are detrapped, the sink strength is much larger than usually used
The sink strength for a defect that is detrapped, is much larger than the values usually used.
%The sink strength is much larger, than is usually used, for defects that are detrapped.
We present a theory how to determine the appropriate sink strength,
and show that the rate equation method, in some cases,
gives wrong results if the detrapping dependence on the sink strength parameter is omitted.
 
\end{abstract}

%\pacs{Valid PACS appear here}% PACS, the Physics and Astronomy
                             % Classification Scheme.
%\keywords{Suggested keywords}%Use showkeys class option if keyword
                              %display desired
\keywords{Mean-field, Rate equations, sink strength, rate theory, defect dynamics, trapping}
\pacs{
   02.30.Jr   %	Partial differential equations
   %02.50.Ng   %	Distribution theory and Monte Carlo studies
   05.       % Statistical physics, thermodynamics, and nonlinear dynamical systems
   05.10.-a   %	Computational methods in statistical physics and nonlinear dynamics
   %05.10.Ln   %	Monte Carlo methods (see also 02.70.Tt, Uu in mathematical methods in physics; for Monte Carlo methods extensively used in subdivisions of physics, see the appropriate section; for example, see 52.65.Pp in plasma simulation)
}

\maketitle

% _____________________________________________________________________________________
% \section{Introduction}

The physical and mechanical properties of materials are largely controlled by their micro structure and defect and impurity concentrations. 
To understand and control these changes during ageing, ion irradiation or annealing,
requires a long time and length scale simulation technique.
The only simulation techniques that are able to fulfil these demanding scales are the mean-field rate equations and kinetic Monte Carlo (KMC) methods.
The KMC is a stochastic simulation method, where all the dynamic properties and reactions for all involved defects have to be known.
The strengths of this method include the ability to take into account expected and unexpected correlated events, e.g. close Frenkel pair annihilation.
However, the time step for the KMC method might be of the order of 10\pot{-11} s with only one self-interstitial atom in the system \cite{Derlet07},
which restricts the accessible time and defect concentrations for this method.

In the mean-field rate equations (RE) all the relevant defect mobilities and reactions are
collected to a set of non-linear differential equations that are solved in time and space
\cite{McNabb63,Wiedersich72,Book_Freeman87,Ahlgren12}.
RE has been extensively used for simulating different dynamic processes in materials.
These studies include finding trapping energies of He and H in vacancies \cite{Baskes83,Myers86},
clustering of irradiation induced vacancies and self-interstitials \cite{Wiedersich72},
He and H bubble formation \cite{Wilson76},
swelling \cite{Brailsford72},
precipitation \cite{Wert49},
fusion edge localized modes simulations \cite{Heinola19},
H isotope exchange \cite{Schmid14,Hodille16,Markelj16} and
simulation of thermal desorption spectrometry (TDS) profiles \cite{Pisarev03,Poon08,Gasparyan15,Hodille16},
to mention just a few.

The dynamic processes and reactions for all involved defects have to be known also for the RE method.
The trapping or annihilation processes are described by the so called sink strength \cite{McNabb63,Brailsford72,Wiedersich72},
and the detrapping or dissociation processes by the detrapping parameter.
The detrapping parameters can be determined by reaction rate theory \cite{Arrhenius1889} and transition state theory \cite{Wigner32,Eyring35}.

The sink strengths have been determined analytically for various symmetric traps including spherical traps, 
dislocation lines and grain boundaries \cite{Wert49,Brailsford81,Wiedersich72,Rouchette14}.
For arbitrarily shaped traps the Monte Carlo (MC) method can to be used \cite{Malerba07,Jansson13,Ahlgren17}.
The sink strength dependence on the trap volume fraction and defect jump length using a
fast MC method is developed in Ref. \cite{Ahlgren17}.
The sink strength also depends on the dimensionality of diffusion \cite{Borodin98,Trinkaus02}.
The effect of the strain field between the defect and trap trap can also be taken into account \cite{Doan03,Ahlgren12}.

The main advantage of the RE method is that it is a very fast method to be used for long time and length scale simulations. 
However, the usefulness of the simulation results depend critically on that the defect interaction 
parameters, i.e. sink strengths are correct.
In this paper, we show that the sink strength actually depends on the detrapping or dissociation process. 
We further show that the RE method, in some cases, gives wrong results if the detrapping dependence on the sink strength
parameter is omitted.

% _______________________________________________________________________________________________
% Theory 1 

The rate equation and sink strength theories are developed in Refs. \cite{Wert49,McNabb63,Brailsford72,Wiedersich72}.
We want to relate the concentration change in time of a diffusing defect $c$ (cm$^{-3}$) due to trapping and detrapping processes.
The simplified rate equation with only one kind of trap and excluding boundary diffusion (constant total $c$ concentration)
and source terms for brevity is given by:
\beqn
   \frac{dc}{dt} &=& - c_e \oint F^{in}dA  +  c_f \oint F^{out}dA  \nonumber \\
                 &=& - D k c + E                                                  \label{Eq_rate_2}
\eeqn
where $c_e$ and $c_f$ are the concentrations of empty and filled traps,
$F^{in}$ and $F^{out}$ are the fluxes of defect $c$ into and out from the traps.
$c_e\oint F^{in}dA$ and $c_f\oint F^{out}dA$ denote the trapped and detrapped defect concentrations per second, where $A$ is the trap surface area.
The trapped defect concentration per second is proportional to the defect concentration $c$,
its diffusion coefficient $D$ (cm$^2$/s) and the {\it sink strength} $k$ (cm$^{-2}$).
$E$ is the thermal emission (detrapping) of defect $c$ from filled trap $c_f$, which from rate theory is:
%$E$ is the thermal emission (detrapping) of defect $c$ from filled trap $c_f$, which from rate theory, \Fig{Fig_detrapping}, is:
\beqn
   E = c_f \nu\ \text{exp}\left( -\frac{E_t}{k_{B}T} \right),
\eeqn
where $\nu$ is the detrapping attempt frequency (Hz), $E_t$ ($\approx E_b+E_m$) is the trapping or dissociation  energy (eV), $E_b$ is the
binding energy and $E_m$ the defect migration barrier, $k_B$ is the Boltzmann constant and $T$ the absolute temperature.

% ________________________________________________________
% MC sink strengths

Usually, the sink strength is determined analytically or by the MC method and the detrapping attempt frequency and the trapping energy
from atomistic simulations or experiments.
The analytical sink strength for spherical traps under 3D diffusion limit with trap radius $R_t$ and
concentration $c_e$ is given by the recursive equation by Brailsford and Bullough \cite{Brailsford81}:
\beqn
  k = 4\pi R_t c_e (1 + R_t \sqrt{k} ).  \label{eq:k2_BB}
\eeqn
%The above equation is derived for spherical cell.
A modified equation taking into account the trap volume fraction and defect jump length is given in \cite{Ahlgren17}.
The sink strengths are defined for randomly distributed defects which results, for instance, from irradiation.
However, if the defects detrap from traps, the initial position is located close to the trap.
We will now show that the sink strength for defects is greatly enhanced if the initial location close to the trap is taken into account.

The sink strengths for spherical traps are simulated by the fast MC method \cite{Ahlgren17}.
The initial position for the defect in the cell is either random or a close distance (0.05 nm) from the trap boundary, as if the
defect would have been detrapped.
The trap and MC parameters are given in \tab{Table:parameters}. The concentrations of the traps was chosen so that
the trap volume fraction is approximately between 5\eex{-6} and 0.2.

\begin{table}[h!b!p!]
%\centering
\caption{The parameters used for MC sink strength simulations. The defect jump length is 0.1 nm and 10\pot{7} defects are simulated
         for each trap concentration giving a statistical error of about 3\eex{-4}\% for the sink strengths \cite{Ahlgren17}.}
\label{Table:parameters}
\begin{tabular}{c c}
%\hline
\hline
Trap radius & Trap concentration    \\
(nm)        & (nm$^{-3}$)             \\
\hline
0.5         & 6\eex{-6} - 0.4       \\        
1.0         & 7\eex{-6} - 5\eex{-2} \\
2.0         & 1\eex{-7} - 6\eex{-3} \\
%Times $\Delta$ saving time ($T_{res}$)  &  Number of jumps saved\\
%\hline
%1000 { } \   & { } 0.01 \% \\
% 100 { } \   & { }1.00 \% \\
%{ } 0.1      &   94.04 \% \\
%{ }{ } 0.01  &   99.19 \% \\
\hline
\end{tabular}\\
%$^a$ Total energy for \Na+1 atoms on \Na lattice sites with 1 interstitial\\
%$\frac{\sum_i^N\Delta R_{MD,i}^2}{6 \sum_i^N \Delta Time_i}$
\end{table}

\begin{figure}[h!]
  \centering
  \includegraphics[width = 10cm]{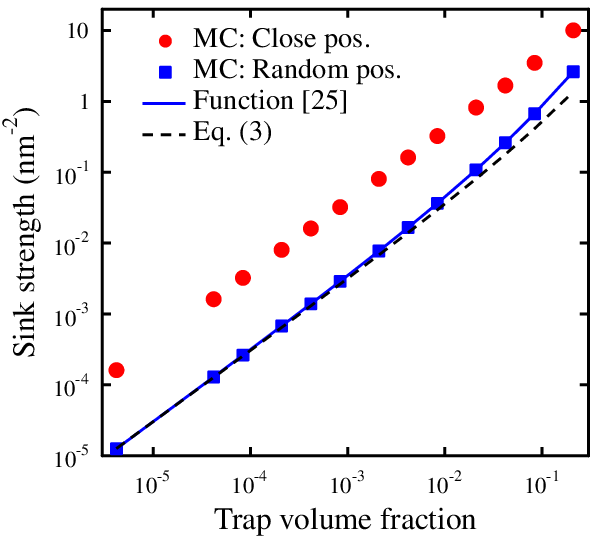}
  \caption{ The sink strengths for spherical trap with trap radius 1.0 nm simulated by MC.
The initial position for the defect in the simulation cell is either random or a close distance
(0.05 nm) from the trap boundary.
The sink strengths for defects with close to trap initial position is greatly enhanced 
compared to the ones with random initial position.
 }
\label{fig:k2_rand_end_close}
\end{figure}

The results for the sink strength simulations are shown in \Fig{fig:k2_rand_end_close}.
The defects with close to trap initial position have a large probability to be trapped
in the trap close to it before diffusing away from it.
Thus, the sink strength is much larger for defects close to the trap than for defects with random initial position. 
Note also that the usually used random sink strength, \eq{eq:k2_BB}, is about 50\% too small
at the highest trap volume fraction.
% while the sink strength equation in Ref. \cite{Ahlgren17} agree with the random MC sink strengths.

% ______________________________________
% Fig 1) Sink strengths close and random 

\begin{figure}[h!]
  \centering
  \includegraphics[width = 10cm]{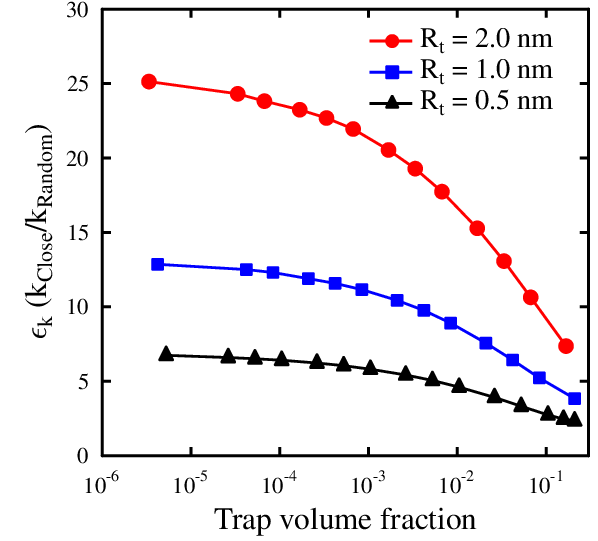}
  \caption{
  The enhancement factor $\epsilon_{k}$ (sink strength for close position divided by the sink strength for random position)
as a function of trap volume fraction for three different trapping radii.
For trapping radius 1.0 nm, the points are from the sink strengths given in \Fig{fig:k2_rand_end_close}.
}
\label{fig:Enhancement}
\end{figure}

\Figure{fig:Enhancement} shows, for three different trapping radii: 0.5, 1.0 and 2.0 nm,
the enhancement factor, which is defined as the sink strength for close position divided by the strength for random position.
The probability to be trapped in the close trap increases with the trap size.
Therefore, the enhancement factor increases with larger traps.
The enhancement factor approaches a maximum value when the trap concentration goes to zero.
For larger trap volume fractions, the trap concentration is larger and the probability 
to be trapped in another trap than the close one increases, leading to a smaller enhancement factor.
% ________________________________________
% Fig 2) Sink strengths enhancement factor 
 
% _______________________________________________________________________________________________
% Theory 2

The sink strengths are clearly larger if the defects are introduced in the cell
from detrapping compared to irradiation.
At equilibrium ($dc/dt=0$) the detrapping and trapping rates are equal and \eq{Eq_rate_2} gives
%\beqn
%   c_e \oint F^{in}dA  = \enhk c_f \oint F^{out}dA,   \label{Eq_equilib_1}
%\eeqn
%which takes into account the enhancement, $\enhk$, due to close detrapping.
%Further at equilibrium
\beqn
   D k c = E =  c_e \oint F^{in}dA = \enhk c_f \oint F^{out}dA,   \label{Eq_equilib_2}
\eeqn
which takes into account the enhancement, $\enhk$, due to close detrapping.
From this we get
%giving with \eq{Eq_equilib_1}
\beqn
   c_f\oint F^{out}dA = E/\enhk,                 \label{Eq_equilib_3}
\eeqn
which inserted in the sink strength solved from \eq{Eq_rate_2} becomes
\beqn
   k  &=& \frac{1}{D\ c}\left[ c_e \oint F^{in}dA  -  c_f \oint F^{out}dA  + E \right] \nonumber\\
      &=& \frac{1}{D\ c}\left[ c_e \oint F^{in}dA  -  E/\enhk + E \right]              \nonumber\\
      &=& \frac{c_e}{D\ c} \oint F^{in}dA  + \frac{E (1-1/\enhk)}{D\ c}                \nonumber\\
      &=& k_{rand} + \frac{E (1-1/\enhk)}{D\ c}.                                                    \label{Eq_k2_last}
\eeqn
The first term is the usual random position sink strength \cite{Brailsford72,Wiedersich72} and the second term 
start enhancing the sink strength when detrapping is activated.
If there is no detrapping or the enhancement factor is one,
the sink strength reduces to the one usually used.
The full rate equation including 3D diffusion and the source of defects (S) from irradiation now reads
\beqn
   \frac{dc}{dt} &=& D\nabla^2c - D k c + E + S,   \label{Eq_rate_Fin1}
\eeqn
where the $k$ is given in \eq{Eq_k2_last}.
An alternative formulation that gives the right equilibrium condition,  % ($D \enhk k_{rand} c = E$),
with the irradiation term omitted,
is obtained if \eq{Eq_k2_last} is inserted in \eq{Eq_rate_Fin1}
\beqn
   \frac{dc}{dt} &=& D\nabla^2c - D k_{rand} c + E/\enhk.   \label{Eq_rate_Fin2}
\eeqn
This alternative form should only be used without the source term.

% _____________________________________________________________________________________________________
% KMC and RE TDS

To check the RE simulation results using different sink strength theories, we do the following test simulation,
where the parameters have been chosen so that KMC simulations are possible to do for comparison.
We have a 200 nm thick layer with a Gaussian trap profile located at the mean depth 50 nm, standard deviation (SD) 10 nm
and the maximum trap concentration value at the mean depth is 1\eex{18} $cm^{-3}$.
Initially the traps are filled with the defects, with one defect per trap.
The different parameters used in the simulations are given in \tab{Table:KMCparameters}.
The $k_{rand}$ is given by \eq{eq:k2_BB}.
The difference in using \eq{eq:k2_BB} or the sink strength from \cite{Ahlgren17} is very small,
because of the low trap volume fraction seen in \Fig{fig:k2_rand_end_close}.
The temperature in the beginning of the simulation is 300 K and increases linearly with 50 K/s to 800 K
during the 10 s simulation.
With increasing temperature in the simulation, defects start to detrap and diffuse in the layer.
Some of them are retrapped but some reach the surface and leave the layer.
This defect surface flux away from the layer is monitored and compared for the RE and KMC simulations.
The simulation corresponds to the popular thermal desorption spectrometry (TDS) method frequently used in
experiments and simulations.
Here, there are no surface parameters, the flux of defects is calculated directly as the
flux ($cm^{-2}s^{-1}$) of atoms crossing the layer boundary. 
In the second simulation we use two different traps to see the effect of traps with different trapping energies.
The defect diffusion parameters for all the simulations are:
jump length 0.1 nm, jump frequency 5\eex{12}Hz and migration barrier 0.25 eV.
The detrapping distance from the traps is 0.05 nm.
Both traps have the Gaussian mean depth 50 nm with standard deviation 10 nm.

\begin{table}[h!b!p!]
%\centering
\caption{The trap parameters used for RE and KMC test simulations.
Only trap 1 is used in the first simulation and both traps in the second one.
The defect detrapping attempt frequency $\nu$ is 5\eex{12} Hz for both traps.
R$_t$ is the trapping radius and E$_b$ the binding energy.
}
\label{Table:KMCparameters}
\begin{tabular}{c c c c}
\hline
%         & \mc{3}{c}{\textbf{Gaussian profile params.}}       \\
Trap     & Gaussian              & R$_t$ & E$_b$   \\
Num.     & Max. conc. ($cm^{-3}$) & (nm)  & (eV)    \\
\hline
\hline
{\bf 1:} & 1\eex{18}  & 1.0   & 0.8     \\
\hline
{\bf 2:} & 5\eex{17}  & 2.0   & 1.0     \\
\hline
\end{tabular}\\
%$^a$ Total energy for \Na+1 atoms on \Na lattice sites with 1 interstitial\\
%$\frac{\sum_i^N\Delta R_{MD,i}^2}{6 \sum_i^N \Delta Time_i}$
\end{table}

\begin{figure}[h!]
  \centering
  \includegraphics[width = 10cm]{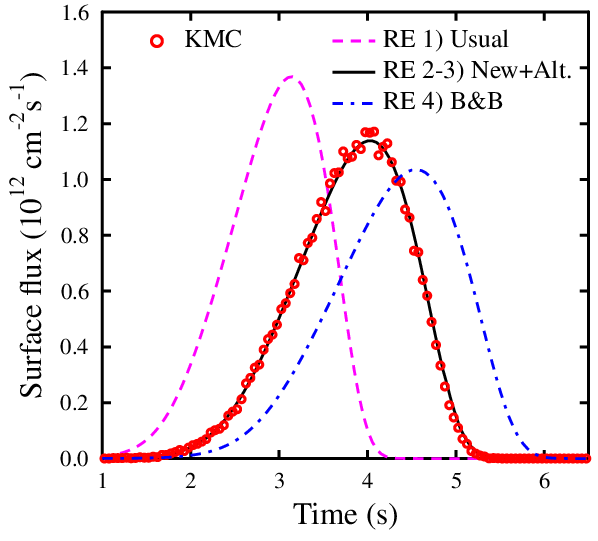}
  \caption{The defect flux comparison between KMC and RE simulations using different sink strength theories (see text).}
  \label{fig:TDS_1}
\end{figure}

\Figure{fig:TDS_1} shows the KMC and RE simulated fluxes of the detrapped and diffusing defects through the front surface.
The RE is now 1D with depth $z$ in the layer and reads
\beqn
   \frac{dc}{dt} &=& D\frac{d^2c}{dz^2} - D k c + E.   \label{Eq_rate_1D}
\eeqn
The different sink strengths and detrapping terms used are given in \tab{Table:k_and_E}.
% ________________________________________________________
% Figure 3) TDS 1
% ________________________________________________________
We can see that the TDS peak (defect flux through the front surface) occurs too fast for the usual RE formalism.
This is to be expected because the sink strength is too small,
it does not take into account the enhanced retrapping due to the defects being detrapped close to the empty trap.
The new theory \eq{Eq_rate_Fin1} and the equivalent formulation \eq{Eq_rate_Fin2} matches nicely with the KMC simulations.
However, even if the TDS peak for \eq{Eq_rate_Fin1} and \eq{Eq_rate_Fin2} are almost identical,
their defect profile in the layer during the simulation is not.
\eq{Eq_rate_Fin1} agrees with the KMC defect profile during the simulation,
detrapping starts similar to KMC and the enhanced trapping delays the defect diffusion
away from the trap profile.
Whereas for \eq{Eq_rate_Fin2} the detrapping is delayed, which in a similar way delays the defect diffusion away from the trap profile.
Using the equations by Brailsford \& Bullough \cite{Brailsford81} gives the TDS peak too late.
%, the fact that the detrapping depends 

\begin{table}[h!b!p!]
%\centering
\caption{The sink strengths and detrapping terms used for the RE test simulations.
         $k_{rand}$ is given by \eq{eq:k2_BB}.
}
\label{Table:k_and_E}
\begin{tabular}{l l l}
\hline
                 & Sink strength,$k$ & Detrapping, $E$     \\
\hline
1) Usual         & $k_{rand}$                             & $c_f \nu\ \text{exp}(-E_t/k_{B}T)$ \\
2) New           & $k_{rand} + E (1-1/\epsilon_k)/(D\ c)$ & $c_f \nu\  \text{exp}(-E_t/k_{B}T)$ \\
3) New (Alt)$^a$ & $k_{rand}$                             & $c_f \nu\  \text{exp}(-E_t/k_{B}T) / \epsilon_k$ \\
4) B\&B$^b$      & $k_{rand}$                             & $4\pi R_t c_f (1 + R_t \sqrt{k}) D \text{exp}(-E_b/k_{B}T)$ \\
\hline
\hline
\end{tabular}\\
%\begin{flushleft}
%$^a$ Alternative formulation \ref{Eq_rate_Fin2}.\hfill  \\
$^a$ Alternative formulation \eq{Eq_rate_Fin2}.  \\
$^b$ Brailsford \& Bullough \cite{Brailsford81}.
%\end{flushleft}
\end{table}

\begin{figure}[h!]
  \centering
  \includegraphics[width = 10cm]{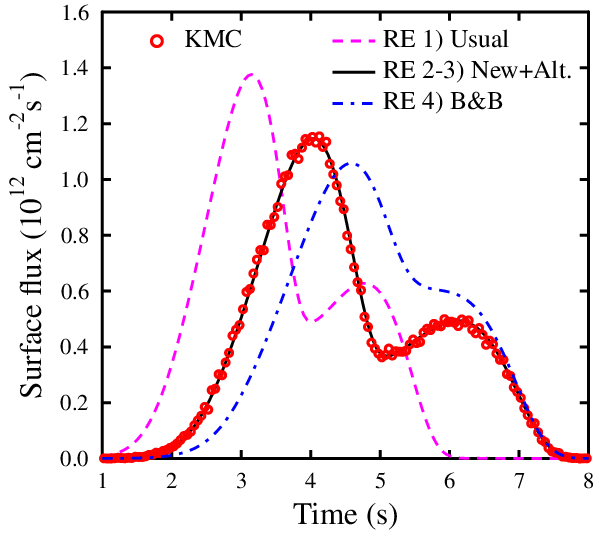}
  \caption{The defect flux comparison between KMC and RE simulations with two traps.}
  \label{fig:TDS_2}
\end{figure}

The simulation with two traps, \Fig{fig:TDS_2}, confirms the results for the one trap simulation.
\eq{Eq_rate_Fin1} and \eq{Eq_rate_Fin2} agree with the KMC simulations, while the TDS peaks are shifted for the other formulations.
The apparent TDS agreement between times 7 and 8 s for the B\&B formulation is a coincidence.
Because, the detrapping term in \tab{Table:k_and_E} with the diffusion coefficient, $D=1/6\cdot 0.1^2 \nu\cdot \text{exp}(-E_m/k_{B}T)$,
approximately becomes:
$4\pi 2 c_f (1/6\cdot 0.1^2\nu ) \text{exp}(-E_t/k_{B}T) \approx  c_f \nu\ \text{exp}(-E_t/k_{B}T) / 24$,
which happens to agree with the alternative formulation \tab{Table:k_and_E}, for the enhancement factor
for trapping radius 2 nm from \Fig{fig:Enhancement}.
The diffusion jump and the detrapping frequencies are the same in these simulations.
It should be noted that even though the simulation was chosen favourable for the KMC technique, the RE
simulation was about 100000 faster.
% ________________________________________________________
% Figure 4) TDS 2

%\section{Discussion}

The results of this study show that the sink strengths can also depend on the detrapping or dissociation process.
Therefore, the numerous simulations including detrapping, done for about 50 years are probably not entirely accurate.
The new sink strength formulation needs the new enhancement factor parameter to be determined.
The enhancement factor can easily be obtained using the MC method, and \Fig{fig:TDS_2} shows that the
factors don't affect each other for trap concentrations below about 1\eex{18} $cm^{-3}$.
However, if the different trap concentrations are higher, then the effect of the other traps on each enhancement factor has to be checked.
Including the detrapping effect in the sink strength interaction parameter will take usefulness of the RE method to a new level.
The RE simulations should now give more reliable defect and trap parameters, and the long length and time scale
micro structure simulations should be more accurate.

% \section{Conclusions}

\section*{Acknowledgements}

This work has been carried out within the framework of the EUROfusion Consortium and has received funding from the 
Euratom research and training programme 2014 - 2018 under grant agreement No 633053.
The views and opinions expressed herein do not necessarily reflect those of the European Commission.
Grants of computer time from the Centre for Scientific Computing in Espoo, Finland, 
are gratefully acknowledged.

% _______________________________________________________________________________________
%\appendix           % /home/tahlgren/home/latex/sink-strengths/error_for_KMC_k2.tex
%\section{Statistical error estimate for sink strengths calculated by the MC method}     \label{app:k2_error}

\section*{References}

%\bibliographystyle{ieeetr}	% https://www.sharelatex.com/learn/Bibtex_bibliography_styles

%\bibliographystyle{apsrev4-1} % Tell bibtex which bibliography style to use

%\bibliography{/home/tahlgren/home/Publications/BibTex-files/simulation,/home/tahlgren/home/Publications/BibTex-files/ownpubs}

\begin{thebibliography}{28}%
\makeatletter
\providecommand \@ifxundefined [1]{%
 \@ifx{#1\undefined}
}%
\providecommand \@ifnum [1]{%
 \ifnum #1\expandafter \@firstoftwo
 \else \expandafter \@secondoftwo
 \fi
}%
\providecommand \@ifx [1]{%
 \ifx #1\expandafter \@firstoftwo
 \else \expandafter \@secondoftwo
 \fi
}%
\providecommand \natexlab [1]{#1}%
\providecommand \enquote  [1]{``#1''}%
\providecommand \bibnamefont  [1]{#1}%
\providecommand \bibfnamefont [1]{#1}%
\providecommand \citenamefont [1]{#1}%
\providecommand \href@noop [0]{\@secondoftwo}%
\providecommand \href [0]{\begingroup \@sanitize@url \@href}%
\providecommand \@href[1]{\@@startlink{#1}\@@href}%
\providecommand \@@href[1]{\endgroup#1\@@endlink}%
\providecommand \@sanitize@url [0]{\catcode `\\12\catcode `\$12\catcode
  `\&12\catcode `\#12\catcode `\^12\catcode `\_12\catcode `\%12\relax}%
\providecommand \@@startlink[1]{}%
\providecommand \@@endlink[0]{}%
\providecommand \url  [0]{\begingroup\@sanitize@url \@url }%
\providecommand \@url [1]{\endgroup\@href {#1}{\urlprefix }}%
\providecommand \urlprefix  [0]{URL }%
\providecommand \Eprint [0]{\href }%
\providecommand \doibase [0]{http://dx.doi.org/}%
\providecommand \selectlanguage [0]{\@gobble}%
\providecommand \bibinfo  [0]{\@secondoftwo}%
\providecommand \bibfield  [0]{\@secondoftwo}%
\providecommand \translation [1]{[#1]}%
\providecommand \BibitemOpen [0]{}%
\providecommand \bibitemStop [0]{}%
\providecommand \bibitemNoStop [0]{.\EOS\space}%
\providecommand \EOS [0]{\spacefactor3000\relax}%
\providecommand \BibitemShut  [1]{\csname bibitem#1\endcsname}%
\let\auto@bib@innerbib\@empty
%</preamble>
\bibitem [{\citenamefont {Derlet}\ \emph {et~al.}(2007)\citenamefont {Derlet},
  \citenamefont {Nguyen-Manh},\ and\ \citenamefont {Dudarev}}]{Derlet07}%
  \BibitemOpen
  \bibfield  {author} {\bibinfo {author} {\bibfnamefont {P.~M.}\ \bibnamefont
  {Derlet}}, \bibinfo {author} {\bibfnamefont {D.}~\bibnamefont {Nguyen-Manh}},
  \ and\ \bibinfo {author} {\bibfnamefont {S.~L.}\ \bibnamefont {Dudarev}},\
  }\href {\doibase 10.1103/PhysRevB.76.054107} {\bibfield  {journal} {\bibinfo
  {journal} {Phys. Rev. B}\ }\textbf {\bibinfo {volume} {76}},\ \bibinfo {eid}
  {054107} (\bibinfo {year} {2007})}\BibitemShut {NoStop}%
\bibitem [{\citenamefont {McNabb}\ and\ \citenamefont
  {Foster}(1963)}]{McNabb63}%
  \BibitemOpen
  \bibfield  {author} {\bibinfo {author} {\bibfnamefont {A.}~\bibnamefont
  {McNabb}}\ and\ \bibinfo {author} {\bibfnamefont {P.~K.}\ \bibnamefont
  {Foster}},\ }\href@noop {} {\bibfield  {journal} {\bibinfo  {journal} {Trans.
  Metal. Soc. AIME}\ }\textbf {\bibinfo {volume} {227}},\ \bibinfo {pages}
  {618} (\bibinfo {year} {1963})}\BibitemShut {NoStop}%
\bibitem [{\citenamefont {Wiedersich}(1972)}]{Wiedersich72}%
  \BibitemOpen
  \bibfield  {author} {\bibinfo {author} {\bibfnamefont {H.}~\bibnamefont
  {Wiedersich}},\ }\href@noop {} {\bibfield  {journal} {\bibinfo  {journal}
  {Radiation Effects}\ }\textbf {\bibinfo {volume} {12}},\ \bibinfo {pages}
  {111} (\bibinfo {year} {1972})}\BibitemShut {NoStop}%
\bibitem [{\citenamefont {Freeman}(1987)}]{Book_Freeman87}%
  \BibitemOpen
  \bibfield  {author} {\bibinfo {author} {\bibfnamefont {G.}~\bibnamefont
  {Freeman}},\ }\href@noop {} {\emph {\bibinfo {title} {Kinetics of
  nonhomogeneous processes, L.K. Mansur: Mechanisms and Kinetics of Radiation
  Effects in Metals and Alloys}}}\ (\bibinfo  {publisher} {John Wiley and Sons,
  New York, NY},\ \bibinfo {year} {1987})\BibitemShut {NoStop}%
\bibitem [{\citenamefont {Ahlgren}\ \emph {et~al.}(2012)\citenamefont
  {Ahlgren}, \citenamefont {Heinola}, \citenamefont {V{\"o}rtler},\ and\
  \citenamefont {Keinonen}}]{Ahlgren12}%
  \BibitemOpen
  \bibfield  {author} {\bibinfo {author} {\bibfnamefont {T.}~\bibnamefont
  {Ahlgren}}, \bibinfo {author} {\bibfnamefont {K.}~\bibnamefont {Heinola}},
  \bibinfo {author} {\bibfnamefont {K.}~\bibnamefont {V{\"o}rtler}}, \ and\
  \bibinfo {author} {\bibfnamefont {J.}~\bibnamefont {Keinonen}},\ }\href
  {\doibase 10.1016/j.jnucmat.2012.04.031} {\bibfield  {journal} {\bibinfo
  {journal} {Journal of Nuclear Materials}\ }\textbf {\bibinfo {volume}
  {427}},\ \bibinfo {pages} {152 } (\bibinfo {year} {2012})}\BibitemShut
  {NoStop}%
\bibitem [{\citenamefont {Baskes}\ and\ \citenamefont
  {Wilson}(1983)}]{Baskes83}%
  \BibitemOpen
  \bibfield  {author} {\bibinfo {author} {\bibfnamefont {M.~I.}\ \bibnamefont
  {Baskes}}\ and\ \bibinfo {author} {\bibfnamefont {W.~D.}\ \bibnamefont
  {Wilson}},\ }\href@noop {} {\bibfield  {journal} {\bibinfo  {journal} {Phys.
  Rev. B}\ }\textbf {\bibinfo {volume} {27}},\ \bibinfo {pages} {2210}
  (\bibinfo {year} {1983})}\BibitemShut {NoStop}%
\bibitem [{\citenamefont {Myers}\ \emph {et~al.}(1983)\citenamefont {Myers},
  \citenamefont {Nordlander}, \citenamefont {Besenbacher},\ and\ \citenamefont
  {N{\o}rskov}}]{Myers86}%
  \BibitemOpen
  \bibfield  {author} {\bibinfo {author} {\bibfnamefont {S.~M.}\ \bibnamefont
  {Myers}}, \bibinfo {author} {\bibfnamefont {P.}~\bibnamefont {Nordlander}},
  \bibinfo {author} {\bibfnamefont {F.}~\bibnamefont {Besenbacher}}, \ and\
  \bibinfo {author} {\bibfnamefont {J.~K.}\ \bibnamefont {N{\o}rskov}},\
  }\href@noop {} {\bibfield  {journal} {\bibinfo  {journal} {Phil. Mag. A}\
  }\textbf {\bibinfo {volume} {48}},\ \bibinfo {pages} {397} (\bibinfo {year}
  {1983})}\BibitemShut {NoStop}%
\bibitem [{\citenamefont {Wilson}\ \emph {et~al.}(1976)\citenamefont {Wilson},
  \citenamefont {Baskes},\ and\ \citenamefont {Bisson}}]{Wilson76}%
  \BibitemOpen
  \bibfield  {author} {\bibinfo {author} {\bibfnamefont {W.~D.}\ \bibnamefont
  {Wilson}}, \bibinfo {author} {\bibfnamefont {M.~I.}\ \bibnamefont {Baskes}},
  \ and\ \bibinfo {author} {\bibfnamefont {C.~L.}\ \bibnamefont {Bisson}},\
  }\href {\doibase 10.1103/PhysRevB.13.2470} {\bibfield  {journal} {\bibinfo
  {journal} {Phys. Rev. B}\ }\textbf {\bibinfo {volume} {13}},\ \bibinfo
  {pages} {2470} (\bibinfo {year} {1976})}\BibitemShut {NoStop}%
\bibitem [{\citenamefont {Brailsford}\ and\ \citenamefont
  {Bullough}(1972)}]{Brailsford72}%
  \BibitemOpen
  \bibfield  {author} {\bibinfo {author} {\bibfnamefont {A.}~\bibnamefont
  {Brailsford}}\ and\ \bibinfo {author} {\bibfnamefont {R.}~\bibnamefont
  {Bullough}},\ }\href {\doibase 10.1016/0022-3115(72)90091-8} {\bibfield
  {journal} {\bibinfo  {journal} {J. Nucl. Mater.}\ }\textbf {\bibinfo {volume}
  {44}},\ \bibinfo {pages} {121 } (\bibinfo {year} {1972})}\BibitemShut
  {NoStop}%
\bibitem [{\citenamefont {Wert}\ and\ \citenamefont {Zener}(1949)}]{Wert49}%
  \BibitemOpen
  \bibfield  {author} {\bibinfo {author} {\bibfnamefont {C.}~\bibnamefont
  {Wert}}\ and\ \bibinfo {author} {\bibfnamefont {C.}~\bibnamefont {Zener}},\
  }\href@noop {} {\bibfield  {journal} {\bibinfo  {journal} {J. Appl. Phys.}\
  }\textbf {\bibinfo {volume} {21}},\ \bibinfo {pages} {5} (\bibinfo {year}
  {1949})}\BibitemShut {NoStop}%
\bibitem [{\citenamefont {Heinola}\ \emph {et~al.}(2019)\citenamefont
  {Heinola}, \citenamefont {Ahlgren}, \citenamefont {Brezinsek}, \citenamefont
  {Vuoriheimo},\ and\ \citenamefont {Wiesen}}]{Heinola19}%
  \BibitemOpen
  \bibfield  {author} {\bibinfo {author} {\bibfnamefont {K.}~\bibnamefont
  {Heinola}}, \bibinfo {author} {\bibfnamefont {T.}~\bibnamefont {Ahlgren}},
  \bibinfo {author} {\bibfnamefont {S.}~\bibnamefont {Brezinsek}}, \bibinfo
  {author} {\bibfnamefont {T.}~\bibnamefont {Vuoriheimo}}, \ and\ \bibinfo
  {author} {\bibfnamefont {S.}~\bibnamefont {Wiesen}},\ }\href {\doibase
  https://doi.org/10.1016/j.nme.2019.03.013} {\bibfield  {journal} {\bibinfo
  {journal} {Nuclear Materials and Energy}\ }\textbf {\bibinfo {volume} {19}},\
  \bibinfo {pages} {397 } (\bibinfo {year} {2019})}\BibitemShut {NoStop}%
\bibitem [{\citenamefont {Schmid}\ \emph {et~al.}(2014)\citenamefont {Schmid},
  \citenamefont {von Toussaint},\ and\ \citenamefont
  {Schwarz-Selinger}}]{Schmid14}%
  \BibitemOpen
  \bibfield  {author} {\bibinfo {author} {\bibfnamefont {K.}~\bibnamefont
  {Schmid}}, \bibinfo {author} {\bibfnamefont {U.}~\bibnamefont {von
  Toussaint}}, \ and\ \bibinfo {author} {\bibfnamefont {T.}~\bibnamefont
  {Schwarz-Selinger}},\ }\href@noop {} {\bibfield  {journal} {\bibinfo
  {journal} {Journal of Applied Physics}\ }\textbf {\bibinfo {volume} {116}},\
  \bibinfo {pages} {134901} (\bibinfo {year} {2014})}\BibitemShut {NoStop}%
\bibitem [{\citenamefont {Hodille}\ \emph {et~al.}(2016)\citenamefont
  {Hodille}, \citenamefont {Ferro}, \citenamefont {Fernandez}, \citenamefont
  {Becquart}, \citenamefont {Angot}, \citenamefont {Layet}, \citenamefont
  {Bisson},\ and\ \citenamefont {Grisolia}}]{Hodille16}%
  \BibitemOpen
  \bibfield  {author} {\bibinfo {author} {\bibfnamefont {E.~A.}\ \bibnamefont
  {Hodille}}, \bibinfo {author} {\bibfnamefont {Y.}~\bibnamefont {Ferro}},
  \bibinfo {author} {\bibfnamefont {N.}~\bibnamefont {Fernandez}}, \bibinfo
  {author} {\bibfnamefont {C.~S.}\ \bibnamefont {Becquart}}, \bibinfo {author}
  {\bibfnamefont {T.}~\bibnamefont {Angot}}, \bibinfo {author} {\bibfnamefont
  {J.~M.}\ \bibnamefont {Layet}}, \bibinfo {author} {\bibfnamefont
  {R.}~\bibnamefont {Bisson}}, \ and\ \bibinfo {author} {\bibfnamefont
  {C.}~\bibnamefont {Grisolia}},\ }\href
  {http://stacks.iop.org/1402-4896/2016/i=T167/a=014011} {\bibfield  {journal}
  {\bibinfo  {journal} {Physica Scripta}\ }\textbf {\bibinfo {volume} {2016}},\
  \bibinfo {pages} {014011} (\bibinfo {year} {2016})}\BibitemShut {NoStop}%
\bibitem [{\citenamefont {Markelj}\ \emph {et~al.}(2016)\citenamefont
  {Markelj}, \citenamefont {Zalo\v{z}nik}, \citenamefont {Schwarz-Selinger},
  \citenamefont {Ogorodnikova}, \citenamefont {Vavpeti\v{c}}, \citenamefont
  {Pelicon},\ and\ \citenamefont {\v{C}ade\v{z}}}]{Markelj16}%
  \BibitemOpen
  \bibfield  {author} {\bibinfo {author} {\bibfnamefont {S.}~\bibnamefont
  {Markelj}}, \bibinfo {author} {\bibfnamefont {A.}~\bibnamefont
  {Zalo\v{z}nik}}, \bibinfo {author} {\bibfnamefont {T.}~\bibnamefont
  {Schwarz-Selinger}}, \bibinfo {author} {\bibfnamefont {O.}~\bibnamefont
  {Ogorodnikova}}, \bibinfo {author} {\bibfnamefont {P.}~\bibnamefont
  {Vavpeti\v{c}}}, \bibinfo {author} {\bibfnamefont {P.}~\bibnamefont
  {Pelicon}}, \ and\ \bibinfo {author} {\bibfnamefont {I.}~\bibnamefont
  {\v{C}ade\v{z}}},\ }\href {\doibase
  https://doi.org/10.1016/j.jnucmat.2015.11.039} {\bibfield  {journal}
  {\bibinfo  {journal} {Journal of Nuclear Materials}\ }\textbf {\bibinfo
  {volume} {469}},\ \bibinfo {pages} {133 } (\bibinfo {year}
  {2016})}\BibitemShut {NoStop}%
\bibitem [{\citenamefont {Pisarev}\ \emph {et~al.}(2003)\citenamefont
  {Pisarev}, \citenamefont {Voskresensky},\ and\ \citenamefont
  {Porfirev}}]{Pisarev03}%
  \BibitemOpen
  \bibfield  {author} {\bibinfo {author} {\bibfnamefont {A.~A.}\ \bibnamefont
  {Pisarev}}, \bibinfo {author} {\bibfnamefont {I.~D.}\ \bibnamefont
  {Voskresensky}}, \ and\ \bibinfo {author} {\bibfnamefont {S.~I.}\
  \bibnamefont {Porfirev}},\ }\href@noop {} {\bibfield  {journal} {\bibinfo
  {journal} {J. Nucl. Mater.}\ }\textbf {\bibinfo {volume} {313--316}},\
  \bibinfo {pages} {604} (\bibinfo {year} {2003})}\BibitemShut {NoStop}%
\bibitem [{\citenamefont {Poon}\ \emph {et~al.}(2008)\citenamefont {Poon},
  \citenamefont {Haasz},\ and\ \citenamefont {Davis}}]{Poon08}%
  \BibitemOpen
  \bibfield  {author} {\bibinfo {author} {\bibfnamefont {M.}~\bibnamefont
  {Poon}}, \bibinfo {author} {\bibfnamefont {A.}~\bibnamefont {Haasz}}, \ and\
  \bibinfo {author} {\bibfnamefont {J.}~\bibnamefont {Davis}},\ }\href@noop {}
  {\bibfield  {journal} {\bibinfo  {journal} {J. Nucl. Mater.}\ }\textbf
  {\bibinfo {volume} {374}},\ \bibinfo {pages} {390 } (\bibinfo {year}
  {2008})}\BibitemShut {NoStop}%
\bibitem [{\citenamefont {Gasparyan}\ \emph {et~al.}(2015)\citenamefont
  {Gasparyan}, \citenamefont {Ogorodnikova}, \citenamefont {Efimov},
  \citenamefont {Mednikov}, \citenamefont {Marenkov}, \citenamefont {Pisarev},
  \citenamefont {Markelj},\ and\ \citenamefont {\v{C}ade\v{z}}}]{Gasparyan15}%
  \BibitemOpen
  \bibfield  {author} {\bibinfo {author} {\bibfnamefont {Y.}~\bibnamefont
  {Gasparyan}}, \bibinfo {author} {\bibfnamefont {O.}~\bibnamefont
  {Ogorodnikova}}, \bibinfo {author} {\bibfnamefont {V.}~\bibnamefont
  {Efimov}}, \bibinfo {author} {\bibfnamefont {A.}~\bibnamefont {Mednikov}},
  \bibinfo {author} {\bibfnamefont {E.}~\bibnamefont {Marenkov}}, \bibinfo
  {author} {\bibfnamefont {A.}~\bibnamefont {Pisarev}}, \bibinfo {author}
  {\bibfnamefont {S.}~\bibnamefont {Markelj}}, \ and\ \bibinfo {author}
  {\bibfnamefont {I.}~\bibnamefont {\v{C}ade\v{z}}},\ }\href {\doibase
  https://doi.org/10.1016/j.jnucmat.2014.11.022} {\bibfield  {journal}
  {\bibinfo  {journal} {Journal of Nuclear Materials}\ }\textbf {\bibinfo
  {volume} {463}},\ \bibinfo {pages} {1013 } (\bibinfo {year}
  {2015})}\BibitemShut {NoStop}%
\bibitem [{\citenamefont {Arrhenius}(1889)}]{Arrhenius1889}%
  \BibitemOpen
  \bibfield  {author} {\bibinfo {author} {\bibfnamefont {S.}~\bibnamefont
  {Arrhenius}},\ }\href@noop {} {\bibfield  {journal} {\bibinfo  {journal} {Z.
  Phys. Chem. (Leipzig)}\ }\textbf {\bibinfo {volume} {4}},\ \bibinfo {pages}
  {226} (\bibinfo {year} {1889})}\BibitemShut {NoStop}%
\bibitem [{\citenamefont {Wigner}(1932)}]{Wigner32}%
  \BibitemOpen
  \bibfield  {author} {\bibinfo {author} {\bibfnamefont {E.}~\bibnamefont
  {Wigner}},\ }\href@noop {} {\bibfield  {journal} {\bibinfo  {journal} {Z.
  Phys. Chem. Abt.}\ }\textbf {\bibinfo {volume} {19}},\ \bibinfo {pages} {203}
  (\bibinfo {year} {1932})}\BibitemShut {NoStop}%
\bibitem [{\citenamefont {Eyring}(1935)}]{Eyring35}%
  \BibitemOpen
  \bibfield  {author} {\bibinfo {author} {\bibfnamefont {H.}~\bibnamefont
  {Eyring}},\ }\href@noop {} {\bibfield  {journal} {\bibinfo  {journal} {J.
  Chem. Phys.}\ }\textbf {\bibinfo {volume} {3}},\ \bibinfo {pages} {107}
  (\bibinfo {year} {1935})}\BibitemShut {NoStop}%
\bibitem [{\citenamefont {Brailsford}\ and\ \citenamefont
  {Bullough}(1981)}]{Brailsford81}%
  \BibitemOpen
  \bibfield  {author} {\bibinfo {author} {\bibfnamefont {A.~D.}\ \bibnamefont
  {Brailsford}}\ and\ \bibinfo {author} {\bibfnamefont {R.}~\bibnamefont
  {Bullough}},\ }\href@noop {} {\bibfield  {journal} {\bibinfo  {journal}
  {Philos. Trans. R. Soc. London}\ }\textbf {\bibinfo {volume} {302}},\
  \bibinfo {pages} {87} (\bibinfo {year} {1981})}\BibitemShut {NoStop}%
\bibitem [{\citenamefont {Rouchette}\ \emph {et~al.}(2014)\citenamefont
  {Rouchette}, \citenamefont {Thuinet}, \citenamefont {Legris}, \citenamefont
  {Ambard},\ and\ \citenamefont {Domain}}]{Rouchette14}%
  \BibitemOpen
  \bibfield  {author} {\bibinfo {author} {\bibfnamefont {H.}~\bibnamefont
  {Rouchette}}, \bibinfo {author} {\bibfnamefont {L.}~\bibnamefont {Thuinet}},
  \bibinfo {author} {\bibfnamefont {A.}~\bibnamefont {Legris}}, \bibinfo
  {author} {\bibfnamefont {A.}~\bibnamefont {Ambard}}, \ and\ \bibinfo {author}
  {\bibfnamefont {C.}~\bibnamefont {Domain}},\ }\href {\doibase
  http://dx.doi.org/10.1016/j.commatsci.2014.02.011} {\bibfield  {journal}
  {\bibinfo  {journal} {Computational Materials Science}\ }\textbf {\bibinfo
  {volume} {88}},\ \bibinfo {pages} {50 } (\bibinfo {year} {2014})}\BibitemShut
  {NoStop}%
\bibitem [{\citenamefont {Malerba}\ \emph {et~al.}(2007)\citenamefont
  {Malerba}, \citenamefont {Becquart},\ and\ \citenamefont
  {Domain}}]{Malerba07}%
  \BibitemOpen
  \bibfield  {author} {\bibinfo {author} {\bibfnamefont {L.}~\bibnamefont
  {Malerba}}, \bibinfo {author} {\bibfnamefont {C.~S.}\ \bibnamefont
  {Becquart}}, \ and\ \bibinfo {author} {\bibfnamefont {C.}~\bibnamefont
  {Domain}},\ }\href@noop {} {\bibfield  {journal} {\bibinfo  {journal} {J.
  Nucl. Mater.}\ }\textbf {\bibinfo {volume} {360}},\ \bibinfo {pages} {159}
  (\bibinfo {year} {2007})}\BibitemShut {NoStop}%
\bibitem [{\citenamefont {Jansson}\ \emph {et~al.}(2013)\citenamefont
  {Jansson}, \citenamefont {Malerba}, \citenamefont {Backer}, \citenamefont
  {Becquart},\ and\ \citenamefont {Domain}}]{Jansson13}%
  \BibitemOpen
  \bibfield  {author} {\bibinfo {author} {\bibfnamefont {V.}~\bibnamefont
  {Jansson}}, \bibinfo {author} {\bibfnamefont {L.}~\bibnamefont {Malerba}},
  \bibinfo {author} {\bibfnamefont {A.~D.}\ \bibnamefont {Backer}}, \bibinfo
  {author} {\bibfnamefont {C.~S.}\ \bibnamefont {Becquart}}, \ and\ \bibinfo
  {author} {\bibfnamefont {C.}~\bibnamefont {Domain}},\ }\href@noop {}
  {\bibfield  {journal} {\bibinfo  {journal} {J. Nucl. Mater.}\ }\textbf
  {\bibinfo {volume} {442}},\ \bibinfo {pages} {218 } (\bibinfo {year}
  {2013})}\BibitemShut {NoStop}%
\bibitem [{\citenamefont {Ahlgren}\ and\ \citenamefont
  {Bukonte}(2017)}]{Ahlgren17}%
  \BibitemOpen
  \bibfield  {author} {\bibinfo {author} {\bibfnamefont {T.}~\bibnamefont
  {Ahlgren}}\ and\ \bibinfo {author} {\bibfnamefont {L.}~\bibnamefont
  {Bukonte}},\ }\href@noop {} {\bibfield  {journal} {\bibinfo  {journal}
  {Journal of Nuclear Materials}\ }\textbf {\bibinfo {volume} {496}},\ \bibinfo
  {pages} {66} (\bibinfo {year} {2017})}\BibitemShut {NoStop}%
\bibitem [{\citenamefont {Borodin}(1998)}]{Borodin98}%
  \BibitemOpen
  \bibfield  {author} {\bibinfo {author} {\bibfnamefont {V.}~\bibnamefont
  {Borodin}},\ }\href {\doibase https://doi.org/10.1016/S0378-4371(98)00338-0}
  {\bibfield  {journal} {\bibinfo  {journal} {Physica A: Statistical Mechanics
  and its Applications}\ }\textbf {\bibinfo {volume} {260}},\ \bibinfo {pages}
  {467 } (\bibinfo {year} {1998})}\BibitemShut {NoStop}%
\bibitem [{\citenamefont {Trinkaus}\ \emph {et~al.}(2002)\citenamefont
  {Trinkaus}, \citenamefont {Heinisch}, \citenamefont {Barashev}, \citenamefont
  {Golubov},\ and\ \citenamefont {Singh}}]{Trinkaus02}%
  \BibitemOpen
  \bibfield  {author} {\bibinfo {author} {\bibfnamefont {H.}~\bibnamefont
  {Trinkaus}}, \bibinfo {author} {\bibfnamefont {H.~L.}\ \bibnamefont
  {Heinisch}}, \bibinfo {author} {\bibfnamefont {A.~V.}\ \bibnamefont
  {Barashev}}, \bibinfo {author} {\bibfnamefont {S.~I.}\ \bibnamefont
  {Golubov}}, \ and\ \bibinfo {author} {\bibfnamefont {B.~N.}\ \bibnamefont
  {Singh}},\ }\href@noop {} {\bibfield  {journal} {\bibinfo  {journal} {Phys.
  Rev. B}\ }\textbf {\bibinfo {volume} {66}},\ \bibinfo {pages} {060105}
  (\bibinfo {year} {2002})}\BibitemShut {NoStop}%
\bibitem [{\citenamefont {Doan}\ and\ \citenamefont {Martin}(2003)}]{Doan03}%
  \BibitemOpen
  \bibfield  {author} {\bibinfo {author} {\bibfnamefont {N.~V.}\ \bibnamefont
  {Doan}}\ and\ \bibinfo {author} {\bibfnamefont {G.}~\bibnamefont {Martin}},\
  }\href {\doibase 10.1103/PhysRevB.67.134107} {\bibfield  {journal} {\bibinfo
  {journal} {Phys. Rev. B}\ }\textbf {\bibinfo {volume} {67}},\ \bibinfo
  {pages} {134107} (\bibinfo {year} {2003})}\BibitemShut {NoStop}%
\end{thebibliography}

%

\end{document}